\documentclass[prd,preprint,superscriptaddress,showpacs,byrevtex]{revtex4}
\usepackage{bm}
\def\fsl#1{\setbox0=\hbox{$#1$}           
   \dimen0=\wd0                                 
   \setbox1=\hbox{/} \dimen1=\wd1               
   \ifdim\dimen0>\dimen1                        
      \rlap{\hbox to \dimen0{\hfil/\hfil}}      
      #1                                        
   \else                                        
      \rlap{\hbox to \dimen1{\hfil$#1$\hfil}}   
      /                                         
   \fi}                                         %
\newcommand{\be}{\begin{equation}}
\newcommand{\ee}{\end{equation}}
\newcommand{\bea}{\begin{eqnarray}}
\newcommand{\eea}{\end{eqnarray}}
\newcommand{\beq}{\begin{equation}}
\newcommand{\eeq}{\end{equation}}
\newcommand{\beqs}{\begin{eqnarray}}
\newcommand{\eeqs}{\end{eqnarray}}

\begin{document}
\title{ Gauge Invariant Noether's Theorem in Yang-Mills Theory  }
\author{Gouranga C Nayak }\thanks{G. C. Nayak was affiliated with C. N. Yang Institute for Theoretical Physics in 2004-2007.}
\affiliation{ C. N. Yang Institute for Theoretical Physics, Stony Brook University, Stony Brook NY, 11794-3840 USA}
\date{\today}
\begin{abstract}
The gauge invariant definition of the spin dependent gluon distribution function from first principle is necessary to study the proton spin crisis at high energy colliders. In this paper we derive the gauge invariant Noether's theorem in Yang-Mills theory by using combined Lorentz transformation plus local non-abelian gauge transformation. We find that the definition of the gauge invariant spin (or orbital) angular momentum of the Yang-Mills field does not exist in Yang-Mills theory although the definition of the gauge invariant spin (or orbital) angular momentum of the quark exists. We show that the gauge invariant definition of the spin angular momentum of the Yang-Mills field in the literature is not correct because of the non-vanishing boundary surface term in Yang-Mills theory. We also find that the Belinfante-Rosenfeld tensor in Yang-Mills theory is not required to obtain the symmetric and gauge invariant energy-momentum tensor of the quark and the Yang-Mills field.
\end{abstract}
\maketitle
\pagestyle{plain}

\pagenumbering{arabic}

\section{Introduction}

The naive parton model predicted that the spin $\frac{1}{2}$ of the proton can be obtained from the spin of the quarks inside the proton \cite{yjaffe,yjaffe1}. This prediction from the naive parton model was ruled out after the EMC collaboration \cite{yemc} observed that the spin of the quarks plus the spin of the antiquarks inside the proton contributes to a very small fraction of the spin $\frac{1}{2}$ of the proton. This disagreement between the naive parton model prediction and the experimental finding is known as the {\it proton spin crisis}.

The {\it proton spin crisis} is still an unsolved problem in particle physics because, at present, all the combined experimental data \cite{yallspin,yrhic,yrhic1,yeic} predicts that the spin of the quarks inside the proton plus the spin of the antiquarks inside the proton plus the spin of the gluons inside the proton contributes to about 50\% of the spin $\frac{1}{2}$ of the proton \cite{ytotals}. The rest of the proton spin may come from other sources such as from the orbital angular momentum of the quarks inside the proton plus the orbital angular momentum of the antiquarks inside the proton and from the orbital angular momentum of the gluons inside the proton \cite{yjaffe,yjaffe1}.

In the parton model the sum of the spin $S_q$ of the quarks inside the proton and the spin $S_{\bar q}$ of the antiquarks inside the proton and the spin $S_g$ of the gluons inside the proton is given by
\bea
S_q+S_{\bar q}+S_g=\frac{1}{2}~\int_0^1 dx ~[\Delta q(x) +\Delta {\bar q}(x) +\Delta g(x)]
\label{ybspn}
\eea
where $\Delta q(x)$, $\Delta {\bar q}(x)$, $\Delta g(x)$ are experimentally measured polarized quark, antiquark, gluon distribution functions respectively inside the proton. In eq. (\ref{ybspn}) the longitudinal momentum fraction of the parton with respect to the proton is given by $x$.

As mentioned above all the combined experimental data \cite{yallspin,yrhic,yrhic1,yeic} predicts \cite{ytotals}
\bea
S_q +S_{\bar q}+ S_g \sim \frac{1}{4}
\label{yaspna}
\eea
which implies that about 50\% of the proton spin is still missing. To remedy this situation it is suggested in the parton model that \cite{yjaffe,yjaffe1}
\bea
\frac{1}{2} = S_q +S_{\bar q}+ S_g + L_q +L_{\bar q} + L_g
\label{yaspn}
\eea
where $L_q$ is the orbital angular momentum of the quarks inside the proton, $L_{\bar q}$ is the orbital angular momentum of the antiquarks inside the proton and $L_g$ is the orbital angular momentum of the gluons inside the proton.

Since the spin of the proton is a physical observable one finds from eq. (\ref{yaspn}) that the $S_q +S_{\bar q}+ S_g + L_q +L_{\bar q} + L_g$ should be gauge invariant. In addition to this, since the polarized parton distribution functions $\Delta q(x)$, $\Delta {\bar q}(x)$, $\Delta g(x)$ are experimentally measured \cite{yallspin,yrhic,yrhic1,yeic} one expects from eq. (\ref{ybspn}) that $S_q$, $S_{\bar q}$, $S_g$ should be gauge invariant.

Note that the definition of the polarized gluon distribution function $\Delta g(x)$ in the parton model in the literature \cite{yjaffe1,ymanh} only corresponds to the definition of the spin angular momentum of the gluon in the light-cone gauge but it does not correspond to the definition of the spin angular momentum of the gluon in any other gauge. Since the polarized gluon distribution function inside the proton is an universal quantity in QCD it should be same in any gauge. This implies that the correct definition of the gauge invariant polarized gluon distribution function inside the proton in QCD from the first principle is missing.

In addition to this, there is a lot of confusion in the literature about the gauge invariant definition of the spin angular momentum of the gluon and about the gauge non-invariant definition of the spin angular momentum of the gluon, see for example, \cite{yjaffe1,yji1,ywata,yhata,ygold}. Hence, in order to study the {\it proton spin crisis}, it is necessary to derive the gauge invariant definition of the angular momentum of the quark and the gauge invariant definition of the angular momentum of the gluon from the first principle.

The first principle method to study conservation of angular momentum in physics is via Noether's theorem by using Lorentz transformation. However, in gauge theory the correct definition of the gauge invariant angular momentum can not be obtained from the Noether's theorem by using Lorentz transformation alone. This is because, in addition to the Lorentz transformation one must also use the local gauge transformation to derive the gauge invariant Noether's theorem in gauge theory to obtain the gauge invariant definition of the angular momentum from the first principle.

Recently we have derived the gauge invariant Noether's theorem in the Dirac-Maxwell theory in \cite{ynk18} and have obtained the correct gauge invariant definition of the angular momentum of the electron and the electromagnetic field. We have shown in \cite{ynk18} that the notion of the gauge invariant definition of the spin angular momentum of the electromagnetic field does not exists and the notion of the gauge invariant definition of the orbital angular momentum of the electromagnetic field does not exists in the Dirac-Maxwell theory. We have also shown that the gauge invariant definition of the spin angular momentum of the electromagnetic field in the literature \cite{yjaffe1,yji1,ywata,yhata,ygold} is not correct because of the non-vanishing boundary surface term in the Dirac-Maxwell theory \cite{ynk18}.

In this paper we will extend this to Yang-Mills theory. We will derive the gauge invariant Noether's theorem in Yang-Mills theory by using combined Lorentz transformation plus local non-abelian gauge transformation in the non-abelian gauge theory. We find that the definition of the gauge invariant spin (or orbital) angular momentum of the Yang-Mills field does not exist in Yang-Mills theory although the definition of the gauge invariant spin (or orbital) angular momentum of the quark exists. We show that the gauge invariant definition of the spin angular momentum of the Yang-Mills field in the literature is not correct because of the non-vanishing boundary surface term in Yang-Mills theory. We also find that the Belinfante-Rosenfeld tensor in Yang-Mills theory is not required to obtain the symmetric and gauge invariant energy-momentum tensor of the quark and the Yang-Mills field.

The paper is organized as follows. In section II we briefly review the gauge non-invariant Noether's theorem in Yang-Mills theory by using Lorentz transformation. In sections III and IV we derive the gauge invariant Noether's theorem of the Yang-Mills field and the the gauge invariant Noether's theorem of the quark respectively in Yang-Mills theory by using combined Lorentz transformation plus local non-abelian gauge transformation. The derivation of the gauge invariant Noether's theorem of the quark plus the Yang-Mills field in Yang-Mills theory is presented in section V. In section VI we show that the symmetric and gauge invariant definition of the energy-momentum tensor of the quark plus the Yang-Mills field in Yang-Mills theory can be obtained without requiring the Belinfante-Rosenfeld tensor. In section VII
we discuss the non-vanishing boundary surface term in Yang-Mills theory and the non-existence of gauge invariant spin and orbital angular momentum of the Yang-Mills field. Section VIII contains conclusions.

\section{ Lorentz Transformation and Gauge Non-Invariant Noether's Theorem in Yang-Mills Theory }

In this section we will briefly review the gauge non-invariant Noether's theorem in Yang-Mills theory which is derived by using Lorentz transformation. Although this is well known in the literature but we will briefly review it here because we will compare it with the gauge invariant Noether's theorem in Yang-Mills theory derived in this paper by using the combined Lorentz transformation plus local non-abelian gauge transformation.

The lagrangian density of the Yang-Mills field is given by
\bea
{\cal L}(x) =-\frac{1}{4}F_{\nu \lambda}^b(x)F^{\nu \lambda b}(x),~~~~~~~~~~~~~~~~F_{\nu \lambda}^b(x) =\partial_\nu A_\lambda^b(x)-\partial_\lambda A_\nu^b(x)+gf^{bcd}A_\nu^c(x)A_\lambda^d(x)
\label{yp5n}
\eea
where $A^{\nu b}(x)$ is the Yang-Mills potential where $\nu =0,1,2,3$ is the Lorentz index and $b=1,...,8$ is the color index. Under infinitesimal Lorentz transformation the space-time coordinate $x^\nu$ transforms as
\bea
x'_\nu=x_\nu + \epsilon_{\nu \lambda}x^\lambda +\Delta_\nu = x_\nu +\delta x_\nu.
\label{ygspn}
\eea
Similarly under infinitesimal Lorentz transformation the functional differential $\delta A_\nu^b(x)$ of the Yang-Mills potential $A_\nu^b(x)$ is given by
\bea
\delta A_\nu^b(x) =-\delta x^\lambda \partial_\lambda A_\nu^b(x)-A_\lambda^b(x)\partial_\nu \delta x^\lambda
\label{yfspn}
\eea
which is not gauge covariant. Using Euler-Lagrangian equation one finds that the Noether's theorem of the Yang-Mills field in Yang-Mills theory derived by using Lorentz transformation is given by
\bea
\partial_\nu [F^{\nu \lambda b}(x) \delta A_\lambda^b(x)] - \delta x^\nu \partial_\nu {\cal L}(x)=0
\label{ys5n1}
\eea
which is not gauge invariant because $\delta A_\nu^b(x)$ in eq. (\ref{yfspn}) is not gauge covariant. In eq. (\ref{ys5n1}) the gauge invariant ${\cal L}(x)$ and the gauge covariant $F_{\nu \lambda}^b(x)$ of the Yang-Mills field are given by eq. (\ref{yp5n}).

Hence one finds that if one uses the Lorentz transformation of the gauge field alone then one ends up getting gauge non-invariant Noether's theorem in Yang-Mills theory given by eq. (\ref{ys5n1}). On the other hand when one uses combined Lorentz transformation plus local gauge transformation of the gauge field then one ends up getting gauge invariant Noether's theorem in Yang-Mills theory, see eq. (\ref{yiat}).

\subsection{ Gauge Non-Invariant Energy-Momentum Tensor of Yang-Mills Field in Yang-Mills Theory From Gauge Non-Invariant Noether's Theorem}

Under space-time translation we find from eq. (\ref{ys5n1}) the continuity equation
\bea
\partial_\nu T^{\nu \lambda}(x)=0
\label{ygnem}
\eea
where
\bea
&&T^{\nu \lambda}(x)= F^{\nu \sigma b}(x) F_\sigma^{~~\lambda b}(x) +\frac{1}{4}g^{\nu \lambda}  F_{\sigma \delta}^b(x)F^{\sigma \delta b}(x) -F^{\nu \sigma b}(x)  D_\sigma[A] A^{\lambda b}(x)
\label{yhnem}
\eea
is the energy-momentum tensor of the Yang-Mills field which is not gauge invariant. In eq. (\ref{yhnem}) the covariant derivative $D^{bc}_\nu[A]$ in the adjoint representation of SU(3) is given by
\bea
D^{bc}_\nu[A]=\delta^{bc} \partial_\nu +gf^{bdc}A_\nu^d(x).
\label{ydaba}
\eea

\subsection{ Gauge Non-Invariant Spin and Orbital Angular Momentum of Yang-Mills Field in Yang-Mills Theory From Gauge Non-Invariant Noether's Theorem}

Under rotation we find from eq. (\ref{ys5n1}) the continuity equation
\bea
\partial_\nu J^{\nu \lambda \delta}(x)=0
\label{yanem}
\eea
where
\bea
J^{0 \nu \lambda}(x)=M^{\nu \lambda}(x)=-F^{0 \lambda b}(x)A^{\nu b}(x)+F^{0 \nu b}(x)A^{\lambda b}(x)- T^{0 \nu}(x) x^\lambda + T^{0 \lambda}(x) x^\nu=S^{\nu \lambda }(x)+L^{\nu \lambda }(x)\nonumber \\
\label{ycnem}
\eea
is the angular momentum tensor of the Yang-Mills field which is not gauge invariant where $T^{\nu \lambda}(x)$ is given by eq. (\ref{yhnem}).

The gauge non-invariant spin angular momentum tensor $S^{\nu \lambda}(x)$ of the Yang-Mills field from eq. (\ref{ycnem}) is given by
\bea
S^{\nu \lambda}(x)=-F^{0 \lambda b}(x)A^{\nu b}(x)+F^{0 \nu b}(x)A^{\lambda b}(x)
\label{ybcnem}
\eea
which gives the gauge non-invariant spin angular momentum ${\vec S}_g$ of the Yang-Mills field
\bea
{\vec S}^{non-inv}_g = \int d^3x ~{\vec E}^b(x)\times {\vec A}^b(x)
\label{ydspn}
\eea
where ${\vec E}^b(x)$ is the chromo-electric field which is gauge covariant and ${\vec A}^b(x)$ is the Yang-Mills vector potential which is not gauge invariant.

The gauge non-invariant orbital angular momentum tensor $L^{\nu \lambda}(x)$ of the Yang-Mills field from eq. (\ref{ycnem}) is given by
\bea
L^{\nu \lambda}(x)=- T^{0 \nu}(x) x^\lambda + T^{0 \lambda}(x) x^\nu
\label{yccnem}
\eea
where $T^{\nu \lambda}(x)$ is given by eq. (\ref{yhnem}) which is not gauge invariant. By using $F_{\nu \lambda}^b(x)$ from eq. (\ref{yp5n}) and $D_\nu^{bc}[A]$ from eq. (\ref{ydaba}) in (\ref{yhnem}) we find
\bea
&&T^{\nu \lambda}(x)= F^{\nu \sigma b}(x) \partial^\lambda A_\sigma^b(x)+\frac{1}{4}g^{\nu \lambda}  F_{\sigma \delta}^b(x)F^{\sigma \delta b}(x).
\label{yhneam}
\eea
Using eq. (\ref{yhneam}) in (\ref{yccnem}) we find that the gauge non-invariant definition of the orbital angular momentum of the Yang-Mills field obtained from the gauge non-invariant Noether's theorem in Yang-Mills theory by using Lorentz transformation alone is given by
\bea
{\vec L}^{non-inv}_g = \int d^3x~E^{jb}(x) ~{\vec r} \times {\vec \nabla} A^{jb}(x).
\label{yhjgk}
\eea

\subsection{ Gauge Non-Invariant Energy-Momentum Tensor of Quark in Yang-Mills Theory From Gauge Non-Invariant Noether's Theorem }

In Yang-Mills theory the lagrangian density ${\cal L}(x)$ of quark is given by
\bea
{\cal L} = \frac{1}{2}{\bar \psi}(x)[i\gamma^\nu ({\overrightarrow \partial}_\nu -igT^b A_\nu^b(x)) -m  ]\psi(x)-\frac{1}{2}{\bar \psi}(x)[i\gamma^\nu ({\overleftarrow \partial}_\nu +igT^b A_\nu^b(x))+m  ]\psi(x)
\label{ykat}
\eea
where $\psi_i(x)$ is the Dirac field of the quark with color index $i=1,2,3$. Note that the suppression of the color index $i$ of the Dirac field $\psi_i(x)$ of the quark is understood in this paper. Under infinitesimal Lorentz transformation the functional differential $\delta \psi(x)$ of the quark field $\psi(x)$ is given by
\bea
&&\delta \psi(x) =\frac{1}{4i}\epsilon_{\nu \lambda} \sigma^{\nu \lambda} \psi(x)-\delta x^\nu \partial_\nu \psi(x) \nonumber \\
&& \delta {\bar \psi}(x) =-\frac{1}{4i}{\bar \psi}(x)\epsilon_{\nu \lambda} \sigma^{\nu \lambda}-\delta x^\nu \partial_\nu {\bar \psi}(x)
\label{ykjt}
\eea
where
\bea
\sigma^{\nu \lambda}=-\frac{1}{2i}[\gamma^\nu,~\gamma^\lambda].
\label{ykit}
\eea
Using Euler-Lagrangian equation one finds that the Noether's theorem of the quark field in Yang-Mills theory derived by using Lorentz transformation is given by
\bea
\partial_\nu \frac{1}{2i} \left[[\delta {\bar \psi}(x)] \gamma^\nu \psi(x)-{\bar \psi}(x)\gamma^\nu \delta \psi(x)\right]=0
\label{ykgt}
\eea
which is not gauge invariant because $\delta \psi(x)$ and $\delta {\bar \psi}(x)$ in eq. (\ref{ykjt}) are not gauge covariant.

Under translation we find from eq. (\ref{ykgt}) the continuity equation
\bea
\partial_\lambda T^{\lambda \nu}(x)=0
\label{ygnel}
\eea
where
\bea
&&T^{\nu \lambda}(x)= -\frac{1}{2i} {\bar \psi}(x)[\gamma^\nu  {\overrightarrow \partial}^\lambda -\gamma^\nu {\overleftarrow \partial}^\lambda ] \psi(x)
\label{yhnel}
\eea
is the energy-momentum tensor of the quark in Yang-Mills theory which is not gauge invariant.

\subsection{ Gauge Non-Invariant Orbital Angular Momentum of Quark in Yang-Mills Theory From Gauge Non-Invariant Noether's Theorem }

Under rotation we find from eq. (\ref{ykgt}) the continuity equation
\bea
\partial_\nu J^{\nu \lambda \delta}(x)=0
\label{yand}
\eea
where
\bea
J^{0 \nu \lambda}(x)=M^{\nu \lambda}(x)=S^{\nu \lambda }(x)+L^{\nu \lambda }(x),~~~~~~~S^{\nu \lambda}(x)=S^{0 \nu \lambda}(x),~~~~~~L^{\nu \lambda}(x)=- T^{0 \nu}(x) x^\lambda + T^{0 \lambda}(x) x^\nu \nonumber \\
\label{ycnd}
\eea
is the angular momentum tensor of the quark in Yang-Mills theory which is not gauge invariant where $T^{\nu \lambda}(x)$ is given by eq. (\ref{yhnel}) and
\bea
4S^{\nu \lambda \delta}(x) = {\bar \psi}(x) \{\gamma^\nu,~\sigma^{\lambda \delta} \} \psi(x).
\label{ykpt}
\eea
From eqs. (\ref{ycnd}) and (\ref{ykpt}) we find that the gauge invariant definition of the spin angular momentum of the quark obtained from the gauge non-invariant Noether's theorem in Yang-Mills theory by using Lorentz transformation alone is given by
\bea
{\vec S}_q = \int d^3x~\psi^\dagger(x) ~{\vec \Sigma}~ \psi(x).
\label{yahjgk}
\eea
Similarly from eqs. (\ref{ycnd}) and (\ref{yhnel}) we find that the gauge non-invariant definition of the orbital angular momentum of the quark obtained from the gauge non-invariant Noether's theorem in Yang-Mills theory by using Lorentz transformation alone is given by
\bea
{\vec L}^{non-inv}_q = \int d^3x~{\vec r} \times [\psi^\dagger(x)[-i{\overrightarrow {\vec \partial}} + i{\overleftarrow {\vec \partial}}]\psi(x)].
\label{yahjgka}
\eea
Note that even if we started with the gauge invariant lagrangian density of the quark in eq. (\ref{ykat}) in Yang-Mills theory we still obtained gauge non-invariant orbital angular momentum of the quark in eq. (\ref{yahjgka}) in Yang-Mills theory. The reason behind this is that we have derived eq. (\ref{yahjgka}) from the Noether's theorem by using the Lorentz transformation alone. On the other hand when we use combined Lorentz transformation plus local non-abelian gauge transformation to derive gauge invariant Noether's theorem in Yang-Mills theory then we will obtain the gauge invariant definition of the angular momentum of the quark, see eq. (\ref{yoquakh}).

\subsection{ Gauge Invariant Angular Momentum of Quark Plus Yang-Mills Field and Lorentz Transformation in Yang-Mills Theory}

In the Yang-Mills theory the Lagrangian density ${\cal L}(x)$ of the quark plus the Yang-Mills field is given by
\bea
&&{\cal L} = \frac{1}{2}{\bar \psi}(x)[i\gamma^\nu ({\overrightarrow \partial}_\nu -igT^b A_\nu^b(x)) -m  ]\psi(x)-\frac{1}{2}{\bar \psi}(x)[i\gamma^\nu ({\overleftarrow \partial}_\nu +igT^b A_\nu^b(x))+m  ]\psi(x)\nonumber \\
&&-\frac{1}{4}F_{\nu \lambda}^b(x)F^{\nu \lambda b}(x).
\label{ylelm}
\eea
Using Euler-Lagrangian equation one finds that the Noether's theorem of the quark and Yang-Mills field in Yang-Mills theory derived by using Lorentz transformation is given by
\bea
\partial_\nu [\frac{1}{2i} [[\delta {\bar \psi}(x)] \gamma^\nu \psi(x)-{\bar \psi}(x)\gamma^\nu \delta \psi(x)]+F^{\nu \lambda b}(x) \delta A_\lambda^b(x)] - \delta x^\nu \partial_\nu {\cal L}(x)=0
\label{ynelm}
\eea
where the lagrangian density ${\cal L}(x)$ is given by eq. (\ref{ylelm}).

Under translation we find from eq. (\ref{ynelm}) the equation
\bea
\partial_\nu [ F^{\nu \delta b}(x) F_\delta^{~~\lambda b}(x) -F^{\nu \delta b}(x)  D_\delta[A] A^{\lambda b}(x) +\frac{1}{4}g^{\nu \lambda}  F_{\delta \sigma }^b(x)F^{\delta \sigma b}(x)+\frac{i}{2} {\bar \psi}(x)[\gamma^\nu {\overrightarrow \partial}^\lambda -\gamma^\nu {\overleftarrow \partial}^\lambda  ] \psi(x)]=0\nonumber \\
\label{yaiet1}
\eea
which gives
\bea
&& \partial_\nu [ F^{\nu \delta b}(x) F_\delta^{~~\lambda b}(x) -\partial_\delta [F^{\nu \delta b}(x)  A^{\lambda b}(x)]+A^{\lambda b}(x) \partial_\delta F^{\nu \delta b}(x)  -F^{\nu \delta b}(x)  gf^{bda}A_\delta^d(x) A^{\lambda a}(x) \nonumber \\
&&+\frac{1}{4}g^{\nu \lambda}  F_{\delta \sigma }^b(x)F^{\delta \sigma b}(x)+\frac{i}{2} {\bar \psi}(x)[\gamma^\nu {\overrightarrow \partial}^\lambda -\gamma^\nu {\overleftarrow \partial}^\lambda  ] \psi(x)]=0.
\label{yaiet2}
\eea
Note from eq. (\ref{yp5n}) that $F_{\nu \lambda }^b(x)$ is antisymmetric in $\nu \leftrightarrow \lambda$. Hence we find from eq. (\ref{yaiet2})
\bea
&& \partial_\nu [ F^{\nu \delta b}(x) F_\delta^{~~\lambda b}(x) +A^{\lambda b}(x) \partial_\delta F^{\nu \delta b}(x)  +A^{\lambda b}(x) gf^{bda}A_\delta^d(x) F^{\nu \delta a}(x)  \nonumber \\
&&+\frac{1}{4}g^{\nu \lambda}  F_{\delta \sigma }^b(x)F^{\delta \sigma b}(x)+\frac{i}{2} {\bar \psi}(x)[\gamma^\nu {\overrightarrow \partial}^\lambda -\gamma^\nu {\overleftarrow \partial}^\lambda  ] \psi(x)]=0.
\label{yaiet}
\eea
The Yang-Mills field equation in the presence of quark is given by \cite{yyang}
\bea
D_\lambda[A] F^{\lambda \nu b}(x)  =g{\bar \psi}(x) T^b\gamma^\nu \psi(x).
\label{ydelt}
\eea
Hence from eqs. (\ref{yaiet}) and (\ref{ydelt}) we find the continuity equation
\bea
\partial_\nu T^{\nu \lambda}(x)=0
\label{ybiet}
\eea
where
\bea
&&T^{\nu \lambda}(x)= F^{\nu \delta b }(x) F_\delta^{~~\lambda b}(x) +\frac{1}{4}g^{\nu \lambda}  F_{\delta \sigma}^b(x)F^{\delta \sigma b}(x)\nonumber \\
&&+\frac{i}{2} {\bar \psi}(x)[\gamma^\nu  ({\overrightarrow \partial}^\lambda -igT^bA^{\lambda b}(x)) -\gamma^\nu ({\overleftarrow \partial}^\lambda +igT^bA^{\lambda b}(x))  ] \psi(x).
\label{ybkxt}
\eea
is the gauge invariant energy-momentum tensor of the quark plus the Yang-Mills field in Yang-Mills theory which is gauge invariant.

Under rotation we find from eq. (\ref{ynelm}) the equation
\bea
&&\partial_\nu [\frac{1}{2} S^{\nu \lambda \delta}(x)+x^\delta [F^{\nu \sigma b}(x) F_\sigma^{~~\lambda b}(x) -F^{\nu \sigma b}(x)  D_\sigma[A] A^{\lambda b}(x) +\frac{1}{4}g^{\nu \lambda}  F_{\mu \sigma}^b(x)F^{\mu \sigma b}(x)\nonumber \\
&&+\frac{i}{2} {\bar \psi}(x)[\gamma^\nu {\overrightarrow \partial}^\lambda -\gamma^\nu {\overleftarrow \partial}^\lambda  ] \psi(x)]-F^{\nu \delta b}(x)A^{\lambda b}(x)]\epsilon_{\lambda \delta} =0.
\label{yfiet}
\eea
From eq. (\ref{yp5n}) we find that $F_{\nu \lambda }^b(x)$ is antisymmetric in $\nu \leftrightarrow \lambda$. Hence we find
\bea
\partial_\nu [x^\delta  F^{\nu \sigma b}(x) \partial_\sigma A^{\lambda b}(x)\epsilon_{\lambda \delta}]
=-\partial_\nu [x^\delta A^{\lambda b}(x) \partial_\sigma F^{\nu \sigma b}(x) \epsilon_{\lambda \delta}] -\partial_\nu [F^{\nu \delta b}(x) A^{\lambda b}(x)\epsilon_{\lambda \delta}].
\label{yhhs}
\eea
Using eq. (\ref{yhhs}) in (\ref{yfiet}) we find
\bea
&&\partial_\nu [\frac{1}{2}S^{\nu \lambda \delta}(x)+x^\delta [F^{\nu \sigma b}(x) F_\sigma^{~~\lambda b}(x) -A^{\lambda b}(x) \partial_\sigma F^{\sigma \nu b }(x)  -A^{\lambda b}(x) gf^{bda} A_\sigma^d(x) F^{\sigma \nu a }(x) \nonumber \\
&&+\frac{1}{4}g^{\nu \lambda}  F_{\mu \sigma}^b(x)F^{\mu \sigma b}(x) +\frac{i}{2} {\bar \psi}(x)[\gamma^\nu {\overrightarrow \partial}^\lambda -\gamma^\nu {\overleftarrow \partial}^\lambda  ] \psi(x)]]\epsilon_{\lambda \delta} =0.
\label{ygiet}
\eea
Using eq. (\ref{ydelt}) in (\ref{ygiet}) we find the continuity equation
\bea
\partial_\nu J^{\nu \lambda \delta}(x)=0
\label{yhiet}
\eea
where
\bea
J^{0 \lambda \delta}(x)=M^{\mu \nu}(x)=S^{0 \lambda \delta}(x)- T^{0 \lambda}(x) x^\delta+ T^{0 \delta}(x) x^\lambda
\label{yjiet}
\eea
is the gauge invariant angular momentum tensor of the quark plus the Yang-Mills field where the gauge invariants $S^{\nu \lambda \delta}(x)$ and $T^{\nu \lambda}(x)$ are given by eqs. (\ref{ykpt}) and (\ref{ybkxt}) respectively.

From eq. (\ref{yjiet}) we find that the gauge invariant angular momentum of the quark plus the Yang-Mills field in the Yang-Mills theory is given by
\bea
&&  {\vec J}_{q+g}=\int d^3x~[ {\vec r} \times [{\vec E}^b(x) \times {\vec B}^b(x)]+ {\vec r} \times [\psi^\dagger(x) [-i{\overrightarrow {\vec D}}+i{\overleftarrow {\vec D}}] \psi(x)]+\psi^\dagger(x) ~{\vec \Sigma}~ \psi(x)] \nonumber \\
\label{yovbtys}
\eea
which is gauge invariant where $D_\nu^{jk}[A]$ is the covariant derivative in the fundamental representation of SU(3) group as given by
\bea
D_\nu^{jk}[A]=\delta^{jk}\partial_\nu -igT^b_{jk}A_\nu^b(x).
\label{ycvf}
\eea

It is useful to mention here that the $T^{\nu \lambda}(x)$ of the Yang-Mills field in eq. (\ref{yhnem})
is not gauge invariant. Similarly the $T^{\nu \lambda}(x)$ of the quark in eq. (\ref{yhnel}) is not gauge invariant. However, the $T^{\mu \nu}(x)$ of the quark plus the Yang-Mills field in eq. (\ref{ybkxt}) is gauge invariant because the gauge non-invariant part $\partial_\mu [F^{\mu \lambda a}(x)  D_\lambda[A] A^\nu(x)]$ in eq. (\ref{yhnem}) and the gauge non-invariant part $\partial_\mu [\frac{i}{2} {\bar \psi}(x)[\gamma^\mu  {\overrightarrow \partial}^\nu -\gamma^\mu {\overleftarrow \partial}^\nu ] \psi(x)]$ in eq. (\ref{yhnel}) combine to give the gauge invariant part $\partial_\mu [\frac{i}{2} {\bar \psi}(x)[\gamma^\mu  ({\overrightarrow \partial}^\nu -igT^aA^{\nu a}(x)) -\gamma^\mu ({\overleftarrow \partial}^\nu +igA^{\nu a}(x))] \psi(x)]$ in eq. (\ref{ybkxt}) [see eqs. (\ref{yaiet}) -(\ref{ybkxt}) for details].

Since the $T^{\mu \nu}(x)$ in eq. (\ref{ybkxt}) is gauge invariant it should also be obtained from the gauge invariant Noether's theorem by using combined Lorentz transformation plus the local gauge transformation, see eq. (\ref{yckxt}). Similarly, since the ${\vec J}_{q+g}$ in eq. (\ref{yovbtys}) is gauge invariant it should also be obtained from the gauge invariant Noether's theorem by using combined Lorentz transformation plus the local gauge transformation, see eq. (\ref{yovbtysa}).

Note that
\bea
&& {\vec J}_{q+g}\neq {\vec L}^{non-inv}_q+{\vec S}_q+{\vec L}^{non-inv}_{g}+{\vec S}^{non-inv}_{g}
\label{yfingti}
\eea
[see eq. (\ref{yfingt}) for the proof] where ${\vec J}_{q+g}$ is the gauge invariant angular momentum of the quark plus the Yang-Mills field as given by eq. (\ref{yovbtys}), ${\vec S}_{q}$ is the gauge invariant spin angular momentum of the quark in Yang-Mills theory as given by eq. (\ref{yahjgk}), ${\vec L}^{non-inv}_{g}$ is the gauge non-invariant orbital angular momentum of the Yang-Mills field as given by eq. (\ref{yhjgk}), ${\vec S}^{non-inv}_{g}$ is the gauge non-invariant spin angular momentum of the Yang-Mills field as given by eq. (\ref{ydspn}) and ${\vec L}^{non-inv}_{q}$ is the gauge non-invariant orbital angular momentum of the quark in Yang-Mills theory as given by eq. (\ref{yahjgka}).

\section{ Gauge Invariant Noether's Theorem in Yang-Mills Theory Using Combined Lorentz Transformation Plus Local Gauge Transformation }\label{yy}

In the previous section we saw that when the Lorentz transformation alone is used to derive the Noether's theorem then we find the gauge non-invariant definition of the spin angular momentum of the Yang-Mills field and the gauge non-invariant definition of the orbital angular momentum of the Yang-Mills field.
In this section we will use combined Lorentz transformation plus local non-abelian gauge transformation to derive gauge invariant Noether's theorem in Yang-Mills theory which will predict the gauge invariant definition of the angular momentum of the Yang-Mills field. In the next section we will derive the gauge invariant definition of the angular momentum of the quark from the gauge invariant Noether's theorem in Yang-Mills theory.

\subsection{ Combined Lorentz Transformation Plus Non-Abelian Local Gauge Transformation in Yang-Mills Theory }

The (infinitesimal) non-abelian local gauge transformation of the Yang-Mills potential $A_\nu^b(x)$ in SU(3) Yang-Mills theory is given by
\bea
A_\nu^{b~~GT}(x)=A_\nu^b(x) +gf^{bda}A_\nu^d(x)\omega^a(x)+\partial_\nu \omega^b(x)=A_\nu^b(x) +D_\nu[A]\omega^b(x)
\label{ye5n}
\eea
where $f^{abc}$ is the structure constant in the adjoint representation of SU(3) group, $\omega^a(x)$ is the space-time dependent gauge transformation parameter and the symbol $GT$ means gauge transformed. Under the non-abelian local gauge transformation of the Yang-Mills potential $A_\nu^b(x)$ as given by eq. (\ref{ye5n}) the Yang-Mills field tensor $F_{\nu \lambda}^b(x)$ in eq. (\ref{yp5n}) transforms gauge covariantly, {\it i. e.}.
\bea
F_{\mu \nu}^{a~~GT}(x) =F_{\mu \nu}^a(x)+gf^{abc}\omega^c(x)F_{\mu \nu}^b(x).
\label{yhnt}
\eea

The Lorentz transformation the Yang-Mills potential $A_\nu^b(x)$ is given by
\bea
A'^b_\nu(x') = \frac{\partial x^\lambda}{\partial x'^\nu} A_\lambda^b(x),~~~~~~~~~~~A'^{\nu b}(x') = \frac{\partial x'^\nu}{\partial x^\lambda} A^{\lambda b}(x)
\label{yaltr}
\eea
and the Lorentz transformation of the Yang-Mills field tensor $F_{\nu \lambda}^b(x)$ is given by
\bea
F'^{\nu \lambda b}(x') =\frac{\partial x'^\nu}{\partial x^\delta} \frac{\partial x'^\lambda}{\partial x^\sigma} F^{\delta \sigma b}(x),~~~~~~~~F'^b_{\nu \lambda}(x') =\frac{\partial x^\delta}{\partial x'^\nu} \frac{\partial x^\sigma}{\partial x'^\lambda} F_{\delta \sigma}^b(x),~~~~~~~F'^{\nu b}_{~~~\lambda}(x') =\frac{\partial x'^\nu}{\partial x^\delta} \frac{\partial x^\sigma}{\partial x'^\lambda} F^{\delta b}_{~~\sigma}(x). \nonumber \\
\label{ybltr}
\eea

Since the Yang-Mills field tensor $F_{\nu \lambda}^b(x)$ is gauge covariant in eq. (\ref{ybltr}) but the the Yang-Mills potential $A_\nu^b(x)$ is not gauge covariant we find that the general transformation of the Yang-Mills potential $A_\nu^b(x)$ is a Lorentz transformation plus local non-abelian gauge transformation given by
\bea
A'^b_\nu(x') =\frac{\partial x^\lambda}{\partial x'^\nu} A_\lambda^b(x) +gf^{baa}A_\nu^d(x')\Lambda^a+ \partial'_\nu \Lambda^b.
\label{yv5nx}
\eea
The functional differential $\delta A_\nu^b(x)$ of the Yang-Mills potential $A_\nu^b(x)$ is given by
\bea
\delta A_\nu^b(x) = A'^b_\nu(x') -A_\nu^b(x) -\delta x^\lambda \partial_\lambda A_\nu^b(x)
\label{ycltr}
\eea
Using eq. (\ref{yv5nx}) in (\ref{ycltr}) and keeping terms up to order $\delta$ we find
\bea
&&\delta A_\nu^b(x) =D_\nu[A] [\Lambda^b- A_\lambda^b(x) \delta x^\lambda]  -\delta x^\lambda F_{\lambda \nu}^b(x)
\label{yhtt}
\eea
where the covariant derivative $D^{bc}_\nu[A]$ in the adjoint representation of SU(3) is given by eq. (\ref{ydaba}) and the Yang-Mills field tensor $F_{\nu \lambda}^b(x)$ is given by eq. (\ref{yp5n}). Making the non-abelian local gauge transformation on eq. (\ref{yhtt}) we find
\bea
&&\delta A_\nu^{b~~GT}(x) =D_\nu^{GT}[A] [\Lambda^{b~~GT}- A_\lambda^{b~~GT}(x) \delta x^\lambda]  -\delta x^\lambda F_{\lambda \nu}^{b~~GT}(x).
\label{yn5nb4}
\eea

\subsection{ Derivation of Gauge Invariant Noether's Theorem in Yang-Mills Theory  }

As mentioned earlier, the Noether's theorem equation in eq. (\ref{ys5n1}) is not gauge invariant because the functional differential $\delta A_\nu^b(x)$ as given by eq. (\ref{yfspn}) is not gauge covariant when the Lorentz transformation is used alone. However, when the combined Lorentz transformation plus local gauge transformation is used then it is possible to obtain the gauge covariant functional differential $\delta A_\nu^b(x)$. For $\delta A_\nu^{n}(x)$ to be gauge covariant we must have
\bea
&&\delta A_\nu^{b~~GT}(x) =\delta A_\nu^{b}(x)+ gf^{bed} \omega^d(x) \delta A_\nu^e(x).
\label{yn5nb3}
\eea
Hence from eqs. (\ref{yhtt}), (\ref{yn5nb3}) and (\ref{yn5nb4}) we find
\bea
&&D_\nu^{GT}[A] [\Lambda^{b~~GT}- A_\lambda^{b~~GT}(x) \delta x^\lambda]  -\delta x^\lambda F_{\lambda \nu}^{b~~GT}(x)\nonumber \\
&&=D_\nu[A] [\Lambda^b- A_\lambda^b(x) \delta x^\lambda]  -\delta x^\lambda F_{\lambda \nu}^b(x)+ gf^{bed} \omega^d(x) [D_\nu[A] [\Lambda^e- A_\lambda^e(x) \delta x^\lambda]  -\delta x^\lambda F_{\lambda \nu}^e(x)]\nonumber \\
\label{yn5nb5}
\eea
which by using eq. (\ref{ye5n}) and (\ref{yhnt}) gives
\bea
&&D_\nu[A][\Lambda^b(A+D[A]\omega)- \Lambda^b(A)- \delta x^\lambda D_\lambda[A]\omega^b(x) ] \nonumber \\
&&+ gf^{bda}[D_\nu[A]\omega^d(x)][\Lambda^a(A+D[A]\omega)-\Lambda^a(A)- \delta x^\lambda D_\lambda[A]\omega^a(x) ]=0
\label{yn5nb7}
\eea
which has a simple solution
\bea
\Lambda^b= A_\lambda^b(x) \delta x^\lambda
\label{yhyt}
\eea
which agrees with \cite{yjackiw,yberg}. From eqs. (\ref{yhyt}) and (\ref{yhtt}) we find
\bea
&&\delta A_\nu^b(x) = -\delta x^\lambda F_{\lambda \nu}^b(x)
\label{yhzt}
\eea
which is gauge covariant.

Hence by using eq. (\ref{yhzt}) in (\ref{ys5n1}) we find
\bea
\partial_\mu [-F^{\mu \nu a}(x) F_{\lambda \nu }^a(x) \delta x^\lambda] -\delta x^\mu \partial_\mu {\cal L}(x)=0
\label{yiat}
\eea
which is the gauge invariant Noether's theorem of the Yang-Mills field in Yang-Mills theory where the Yang-Mills field lagrangian density ${\cal L}(x)$ is given by eq. (\ref{yp5n}).

\subsection{  Gauge Invariant Energy-Momentum Tensor of Yang-Mills Field From Gauge Invariant Noether's Theorem in Yang-Mills Theory}

Under space-time translation we find from eq. (\ref{yiat}) the continuity equation
\bea
\partial_\nu T^{\nu \lambda}(x)=0
\label{yiet}
\eea
where
\bea
T^{\nu \lambda}(x)= F^{\nu \delta b}(x) F_\delta^{~~\lambda b}(x)+\frac{1}{4}g^{\nu \lambda}  F_{\delta \sigma}^b(x)F^{\delta \sigma b}(x)
\label{yigt}
\eea
is the gauge invariant energy-momentum tensor of the Yang-Mills field in Yang-Mills theory. In eq. (\ref{yigt}) the non-abelian Yang-Mills field tensor $F_{\nu \lambda}^b(x)$ is given by eq. (\ref{yp5n}).

Note that the gauge invariant energy-momentum tensor $T^{\nu \lambda}(x)$ of the Yang-Mills field in eq. (\ref{yigt}) is also symmetric. Hence we find that the Belinfante tensor is not necessary to make the energy-momentum tensor of the Yang-Mills field in eq. (\ref{yigt}) symmetric. The gauge invariant and symmetric energy-momentum tensor $T^{\nu \lambda}(x)$ of the Yang-Mills field in eq. (\ref{yigt}) is obtained from the first principle by using the combined Lorentz transformation plus non-abelian local gauge transformation without requiring the Belinfante tensor.

\subsection{  Gauge Invariant Angular Momentum of Yang-Mills Field From Gauge Invariant Noether's Theorem in Yang-Mills Theory}

Under rotation we find from eq. (\ref{yiat})
\bea
\partial_\nu [ T^{\nu \lambda}(x) x^\delta] \epsilon_{\lambda \delta}=0
\label{yjmt}
\eea
where the gauge invariant and symmetric energy-momentum tensor $T^{\nu \lambda}(x)$ of the Yang-Mills field in Yang-Mills theory is given by eq. (\ref{yigt}).
Note that $\epsilon_{\nu \lambda}$ is antisymmetric, {\it i. e.},
\bea
\epsilon_{\nu \lambda}=-\epsilon_{\lambda \nu}.
\label{yanst}
\eea
Hence using eq. (\ref{yanst}) in (\ref{yjmt}) we find the continuity equation
\bea
\partial_\nu J^{\nu \lambda \delta}(x)=0
\label{yjpt}
\eea
where
\bea
J^{0 \nu \lambda}(x)=M^{\nu \lambda }(x)=T^{0 \lambda}(x) x^\nu - T^{0 \nu}(x) x^\lambda
\label{ykkqt}
\eea
is the angular momentum tensor of the Yang-Mills field which is gauge invariant because $T^{\mu \nu}(x)$ is gauge invariant, see eq. (\ref{yigt}).

From eq. (\ref{ykkqt}) we find that the gauge invariant definition of the angular momentum of the Yang-Mills field obtained from the gauge invariant Noether's theorem by using combined Lorentz transformation plus non-abelian local gauge transformation in the Yang-Mills theory is given by
\bea
&&  {\vec J}_{g}=\int d^3x~ {\vec r} \times [{\vec E}^b(x) \times {\vec B}^b(x)].
\label{yohav}
\eea

\section{ Gauge Invariant Noether's Theorem of Quark in Yang-Mills Theory }\label{yq}

In the previous section we have derived the gauge invariant Noether's theorem of Yang-Mills field by using combined Lorentz transformation plus local non-abelian gauge transformation in Yang-Mills theory and have obtained the gauge invariant definition of the angular momentum of the Yang-Mills field. In this section we will extend this to quark in Yang-Mills theory, {\it i. e.}, we will derive the gauge invariant Noether's theorem of the quark by using combined Lorentz transformation plus local non-abelian gauge transformation in Yang-Mills theory and will obtain the gauge invariant definition of the angular momentum of the quark.

\subsection{ Combined Lorentz Transformation Plus Local Non-Abelian Gauge Transformation of Dirac Field of Quark in Yang-Mills Theory }

The lagrangian density ${\cal L}(x)$ of the quark in the Yang-Mills theory is given by eq. (\ref{ykat}). The functional differential $\delta \psi(x)$ of the Dirac spinor $\psi(x)$ of the quark in Yang-Mills theory is given by
\bea
\delta \psi(x) = \psi'(x') -\psi(x) -\delta x^\lambda \partial_\lambda \psi(x)
\label{ycltrf}
\eea

The infinitesimal local non-abelian gauge transformation of the Dirac field $\psi(x)$ of the quark is given by
\bea
\psi^{GT}(x)=\psi(x) +i gT^b \omega^b(x) \psi(x).
\label{ykrt}
\eea
The lagrangian density ${\cal L}(x)$ of the quark in the Yang-Mills theory in eq. (\ref{ykat}) is gauge invariant under the local non-abelian gauge transformations of the Yang-Mills field $A_\nu^b(x)$ and the quark field $\psi(x)$ as given by eqs. (\ref{ye5n}) and (\ref{ykrt}) respectively.

Since the lagrangian density ${\cal L}(x)$ of the quark in the Yang-Mills theory in eq. (\ref{ykat}) is gauge invariant we find from eq. (\ref{yv5nx}) that the combined Lorentz transformation plus local non-abelian gauge transformation of the Dirac field $\psi(x)$ of the quark in Yang-Mills theory is given by
\bea
&& \psi'(x')=\psi(x)+\frac{1}{4i}\epsilon_{\nu \lambda} \sigma^{\nu \lambda} \psi(x)+igT^b \Lambda^b \psi(x) \nonumber \\
&& {\bar \psi}'(x')={\bar \psi}(x)-\frac{1}{4i}{\bar \psi}(x)\epsilon_{\nu \lambda}\sigma^{\nu \lambda}- {\bar \psi(x)} igT^b \Lambda^b.
\label{ykst}
\eea

\subsection{ Derivation of Gauge Invariant Noether's Theorem of Quark in Yang-Mills Theory  }

For $\Lambda^b$ given by eq. (\ref{yhyt}) the functional differential of the Yang-Mills field $\delta A_\nu^b(x)$ is gauge covariant in eq. (\ref{yhzt}). From eqs. (\ref{ykst}), (\ref{yhyt}) and (\ref{ycltrf}) we find
\bea
&&\delta \psi(x) =\frac{1}{4i}\epsilon_{\nu \lambda} \sigma^{\nu \lambda} \psi(x)-(\delta x^\nu) ({\overrightarrow \partial}_\nu -igT^bA_\nu^b(x)) \psi(x) \nonumber \\
&& \delta {\bar \psi}(x) =-\frac{1}{4i}{\bar \psi}(x)\epsilon_{\nu \lambda}\sigma^{\nu \lambda}- {\bar \psi}(x) ({\overleftarrow \partial_\nu}+igT^bA_\nu^b(x))\delta x^\nu
\label{yktt}
\eea
which is gauge covariant.

Hence by using eqs. (\ref{yktt}) in (\ref{ykgt}) we find that the gauge invariant Noether's theorem of quark in Yang-Mills theory under combined Lorentz transformation plus local non-abelian gauge transformation is given by
\bea
\partial_\nu [\frac{1}{2} S^{\nu \lambda \delta}(x) \epsilon_{\lambda \delta} -\frac{i}{2}(\delta x_\lambda) {\bar \psi}(x)\gamma^\nu [{\overrightarrow \partial}^\lambda -igT^bA^{\lambda b}(x)] \psi(x)+
 \frac{i}{2}{\bar \psi}(x) [{\overleftarrow \partial^\lambda}+igT^bA^{\lambda b}(x)] \gamma^\nu \psi(x) \delta x_\lambda ]=0\nonumber \\
\label{ykut}
\eea
where $S^{\mu \nu \lambda}(x)$ is given by eq. (\ref{ykpt}) which is gauge invariant.

\subsection{  Gauge Invariant Energy-Momentum Tensor of Quark From Gauge Invariant Noether's Theorem in Yang-Mills Theory}

Under translation we find from eq. (\ref{ykut}) the continuity equation
\bea
\partial_\nu T^{\nu \lambda}(x) = 0
\label{ykwt}
\eea
where
\bea
T^{\nu \lambda}(x)= \frac{i}{2} {\bar \psi}(x)[\gamma^\nu  ({\overrightarrow \partial}^\lambda -igT^bA^{\lambda b}(x)) -\gamma^\nu ({\overleftarrow \partial}^\lambda +igT^bA^{\lambda b}(x))  ] \psi(x)
\label{ykxt}
\eea
is the energy-momentum tensor of the quark in Yang-Mills theory which is gauge invariant.

From eq. (\ref{ykxt}) we find that the gauge invariant definition of the energy-momentum tensor $T^{\nu \lambda}(x)$ of the quark in Yang-Mills theory is obtained from the gauge invariant Noether's theorem by using combined Lorentz transformation plus local non-abelian gauge transformation.

\subsection{  Gauge Invariant Angular Momentum of Quark From Gauge Invariant Noether's Theorem in Yang-Mills Theory}

Under rotation we find from eq. (\ref{ykut}) the continuity equation
\bea
\partial_\nu J^{\nu \lambda \delta}(x)=0
\label{ypat}
\eea
where
\bea
J^{0 \nu \lambda}(x)=M^{\nu \lambda}(x)=S^{0 \nu \lambda}(x)- T^{0 \nu}(x) x^\lambda + T^{0 \lambda}(x) x^\nu
\label{ypbtm}
\eea
is the gauge invariant angular momentum tensor of the quark in Yang-Mills theory where the gauge invariants $S^{\nu \lambda \delta}(x)$ and $T^{\nu \lambda}(x)$ are given by eqs. (\ref{ykpt}) and (\ref{ykxt}) respectively.

From eq. (\ref{ypbtm}) we find that the gauge invariant definition of the spin angular momentum of the quark in Yang-Mills theory is given by
\bea
&&  {\vec S}_{q}=\int d^3x~\psi^\dagger(x) ~{\vec \Sigma}~ \psi(x)
\label{yoqua}
\eea
and the gauge invariant definition of the orbital angular momentum of the quark in Yang-Mills theory is given by
\bea
&&  {\vec L}_{q}=\int d^3x~[{\vec r} \times [\psi^\dagger(x) [-i{\overrightarrow {\vec D}}+i{\overleftarrow {\vec D}}] \psi(x)]]
\label{yoquakh}
\eea
where $D_\nu^{jk}[A]$ is the covariant derivative in the fundamental representation of SU(3) group as given by eq. (\ref{ycvf}).

Hence we find that, unlike the gauge non-invariant orbital angular momentum ${\vec L}^{non-inv}_{q}$ in eq. (\ref{yahjgka}) which was derived from gauge non-invariant Noether's theorem by using Lorentz transformation alone, the gauge invariant definition of the orbital angular momentum ${\vec L}_q$ of the quark in eq. (\ref{yoquakh}) in Yang-Mills theory is derived from the gauge invariant Noether's theorem by using combined Lorentz transformation plus local non-abelian gauge transformation.

\section{ Gauge Invariant Noether's Theorem of Quark Plus Yang-Mills Field in Yang-Mills Theory  }

In the section \ref{yy} we have derived the gauge invariant Noether's theorem of Yang-Mills field by using combined Lorentz transformation plus local non-abelian gauge transformation in Yang-Mills theory and have obtained the gauge invariant definition of the angular momentum of the Yang-Mills field. Similarly in the section \ref{yq} we have derived the gauge invariant Noether's theorem of the quark by using combined Lorentz transformation plus local non-abelian gauge transformation in Yang-Mills theory and have obtained the gauge invariant definition of the angular momentum of the quark. In this section we will derive the gauge invariant Noether's theorem of the quark plus the Yang-Mills field by using combined Lorentz transformation plus local non-abelian gauge transformation in Yang-Mills theory and will obtain the gauge invariant definition of the angular momentum of the quark plus the Yang-Mills field.

The gauge invariant lagrangian density of the quark plus the Yang-Mills field in Yang-Mills theory is given by eq. (\ref{ylelm}). Under combined Lorentz transformation plus local non-abelian gauge transformation in Yang-Mills theory the gauge covariant functional differential of the Yang-Mills field $\delta A_\nu^b(x)$ is given by eq. (\ref{yhzt}) and the gauge covariant functional differential of the quark field $\delta \psi(x)$ is given by eq. (\ref{yktt}).

Hence by using eqs. (\ref{yhzt}) and (\ref{yktt}) in eq. (\ref{ynelm}) we find that the gauge invariant Noether's theorem of the quark plus the Yang-Mills field in Yang-Mills theory is given by
\bea
&&\partial_\nu [\frac{1}{2} S^{\nu \lambda \delta}(x) \epsilon_{ \lambda \delta} -\frac{i}{2}(\delta x_\lambda) {\bar \psi}(x)\gamma^\nu [{\overrightarrow \partial}^\lambda -igT^bA^{\lambda b}(x)] \psi(x)+ \frac{i}{2}{\bar \psi}(x) [{\overleftarrow \partial^\lambda}+igT^bA^{\lambda b}(x)] \gamma^\nu \psi(x) \delta x_\lambda\nonumber \\
&&-F^{\nu \lambda b}(x) F_{\delta \lambda}^b(x) \delta x^\delta] -\delta x^\nu \partial_\nu {\cal L}(x)=0
\label{yntdm}
\eea
where the gauge invariant lagrangian density ${\cal L}(x)$ of the quark plus the Yang-Mills field is given by eq. (\ref{ylelm}).

\subsection{  Gauge Invariant Energy-Momentum Tensor of Quark Plus Yang-Mills Field From Gauge Invariant Noether's Theorem in Yang-Mills Theory}

Under translation we find from eq. (\ref{yntdm}) the continuity equation
\bea
\partial_\nu T^{\nu \lambda}(x)=0
\label{ytmdm}
\eea
where
\bea
&&T^{\nu \lambda}(x)= F^{\nu \delta b}(x) F_\delta^{~~\lambda b}(x) +\frac{1}{4}g^{\nu \lambda}  F_{\delta \sigma}^b(x)F^{\delta \sigma b}(x)\nonumber \\
&&+\frac{i}{2} {\bar \psi}(x)[\gamma^\nu  ({\overrightarrow \partial}^\lambda -igT^bA^{\lambda b}(x)) -\gamma^\nu ({\overleftarrow \partial}^\lambda +igT^bA^{\lambda b}(x))  ] \psi(x).
\label{yckxt}
\eea
is the energy-momentum tensor of the quark plus the Yang-Mills field in Yang-Mills theory which is gauge invariant.

From eq. (\ref{yckxt}) we find that the gauge invariant definition of the energy-momentum tensor $T^{\nu \lambda}(x)$ of the quark plus the Yang-Mills field in Yang-Mills theory is obtained from the gauge invariant Noether's theorem by using combined Lorentz transformation plus local non-abelian gauge transformation.

\subsection{  Gauge Invariant Angular Momentum of Quark Plus Yang-Mills Field From Gauge Invariant Noether's Theorem in Yang-Mills Theory}

Under rotation we find from eq. (\ref{yntdm}) the continuity equation
\bea
\partial_\nu J^{\nu \lambda \delta}(x)=0
\label{ygieem}
\eea
where
\bea
J^{ \nu \lambda \delta }(x)=S^{\nu \lambda \delta}(x)- T^{\nu \lambda}(x) x^\delta + T^{\nu \delta}(x) x^\lambda
\label{yfieem}
\eea
which gives
\bea
M^{\lambda \delta }(x)=J^{0 \lambda \delta }(x)=S^{0 \lambda \delta}(x)- T^{0 \lambda}(x) x^\delta + T^{0 \delta}(x) x^\lambda
\label{yhieem}
\eea
which is the gauge invariant angular momentum tensor of the quark plus the Yang-Mills field in Yang-Mills theory where the gauge invariants $S^{\nu \lambda \delta}(x)$ and $T^{\nu \lambda}(x)$ are given by eqs. (\ref{ykpt}) and (\ref{yckxt}) respectively.

From eq. (\ref{yfieem}) we find that the gauge invariant definition of the angular momentum of the quark plus the Yang-Mills field in the Yang-Mills theory is given by
\bea
&&  {\vec J}_{q+g}=\int d^3x~[ {\vec r} \times [{\vec E}^b(x) \times {\vec B}^b(x)]+ {\vec r} \times [\psi^\dagger(x) [-i{\overrightarrow {\vec D}}+i{\overleftarrow {\vec D}}] \psi(x)]+\psi^\dagger(x) ~{\vec \Sigma}~ \psi(x)] \nonumber \\
&& = {\vec J}_{g}+{\vec L}_q+{\vec S}_q
\label{yovbtysa}
\eea
where ${\vec J}_{g}$ is the gauge invariant angular momentum of the Yang-Mills field as given by eq. (\ref{yohav}), ${\vec L}_{q}$ is the gauge invariant orbital angular momentum of the quark in Yang-Mills theory as given by eq. (\ref{yoquakh}) and ${\vec S}_{q}$ is the gauge invariant spin angular momentum of the quark in Yang-Mills theory as given by eq. (\ref{yoqua}).

From eq. (\ref{yovbtysa}) we find that the gauge invariant definition of the angular momentum ${\vec J}_{q+g}$ of the quark plus the Yang-Mills field in Yang-Mills theory is obtained from the gauge invariant Noether's theorem by using combined Lorentz transformation plus local non-abelian gauge transformation.

\section{ Symmetric and Gauge Invariant Energy-Momentum Tensor of Quark Plus Yang-Mills Field in Yang-Mills Theory Without Belinfante-Rosenfeld Tensor}

Note that the energy momentum-tensor $T^{\nu \lambda}(x)$ of the quark plus the Yang-Mills field in eq. (\ref{yckxt}) in Yang-Mills theory is gauge invariant but is not symmetric. First of all note from eqs. (\ref{ygieem}) and (\ref{yfieem}) that
\bea
\partial_\nu [T^{\nu \delta}(x) x^\lambda - T^{\nu \lambda}(x) x^\delta] \neq 0,~~~~~~~~~~~~~{\rm but}~~~~~~~~~~~~~\partial_\nu J^{\nu \lambda \delta}(x)=0
\label{ytbtm}
\eea
which implies that the gauge invariant energy-momentum tensor $T^{\nu \lambda}(x)$ of the quark plus the Yang-Mills field in eq. (\ref{yckxt}) in Yang-Mills theory is not required to be symmetric to obtain the continuity equation of the 3-rd rank total angular momentum momentum tensor $J^{\nu \lambda \delta}(x)$ of the quark plus the Yang-Mills field in Yang-Mills theory.

It is often argued in the literature that the Belinfante-Rosenfeld tensor is required to make the energy momentum-tensor symmetric. However, we will show in this section that the gauge invariant and symmetric energy momentum-tensor of the quark plus the Yang-Mills field can be obtained in Yang-Mills theory without requiring the Belinfante-Rosenfeld tensor. This can be shown as follows.

From the gauge invariant lagrangian density of the quark plus the Yang-Mills field in eq. (\ref{ylelm}) we find by using the
Euler-Lagrange equation in Yang-Mills theory the following Dirac equation for the quark field
\bea
&&i\gamma^\delta {\overrightarrow \partial}_\delta \psi(x) = [m -gT^b\gamma^\delta A_\delta^b (x)]\psi(x), \nonumber \\
&& {\bar \psi}(x){\overleftarrow \partial}_\delta  i\gamma^\delta =-{\bar \psi}(x)[ m -gT^b\gamma^\delta A_\delta^b (x)].
\label{yddrl}
\eea
Using the eq. (\ref{yddrl}) in (\ref{yckxt}) and utilizing the algebra of the Dirac matrices we find
\bea
T^{\lambda \nu}(x)-T^{\nu \lambda}(x) =\partial_\delta [\frac{1}{4}{\bar \psi}(x) \{\gamma^\delta,~\sigma^{\nu \lambda } \} \psi(x)].
\label{yderl}
\eea
Simplifying the algebra of the Dirac matrices we obtain the following antisymmetric relations
\bea
&&\{\gamma^\delta, \sigma^{\nu \lambda}\}=-\{\gamma^\delta, \sigma^{\lambda \nu}\},~~~~~~~~~\{\gamma^\delta, \sigma^{\nu \lambda}\}=-\{\gamma^\lambda, \sigma^{\nu \delta }\},~~~~~~~~~~\{\gamma^\delta, \sigma^{\nu \lambda}\}=-\{\gamma^\nu, \sigma^{\delta \lambda}\}.\nonumber \\
\label{yedrl}
\eea
From eq. (\ref{yedrl}) we find that the 3-rd rank spin angular momentum tensor ${\bar \psi}(x) \{\gamma^\delta,~\sigma^{\nu \lambda } \} \psi(x)$ of the quark is antisymmetric with respect to the indices $\delta \leftrightarrow \nu$, $\delta \leftrightarrow \lambda$  and $\nu \leftrightarrow \lambda$ which implies that
\bea
\partial_\nu \partial_\delta [{\bar \psi}(x) \{\gamma^\delta,~\sigma^{\nu \lambda } \} \psi(x)]
=\partial_\lambda \partial_\delta [{\bar \psi}(x) \{\gamma^\delta,~\sigma^{\nu \lambda } \} \psi(x)]
=\partial_\nu \partial_\lambda [{\bar \psi}(x) \{\gamma^\delta,~\sigma^{\nu \lambda } \} \psi(x)]=0.\nonumber \\
\label{ydfrl}
\eea
By writing eq. (\ref{ytmdm}) as
\bea
\partial_\nu T^{\nu \lambda}(x)=\partial_\nu [\frac{1}{2}T^{\nu \lambda}(x) +\frac{1}{2}T^{\lambda \nu}(x)+\frac{1}{2}T^{\nu \lambda}(x)-\frac{1}{2}T^{\lambda \nu}(x)]= 0
\label{yakwt}
\eea
and then utilizing eqs. (\ref{yderl}), (\ref{ydfrl}) and (\ref{yckxt}) in eq. (\ref{yakwt}) we find the continuity equation
\bea
\partial_\nu T_S^{\nu \lambda}(x) =0
\label{yakwtz}
\eea
where $T_S^{\mu \nu}(x)$ is the symmetric and gauge invariant energy-momentum tensor of the quark plus the Yang-Mills field in Yang-Mills theory which is given by
\bea
&&T_S^{\nu \lambda}(x)= F^{\nu \delta b}(x) F_\delta^{~~\lambda b}(x) +\frac{1}{4}g^{\nu \lambda}  F_{\delta \sigma}^b(x)F^{\delta \sigma b}(x)\nonumber \\
&&+\frac{i}{4} {\bar \psi}(x)[\gamma^\nu  ({\overrightarrow \partial}^\lambda -igT^bA^{\lambda b}(x)) +\gamma^\lambda  ({\overrightarrow \partial}^\nu -igT^bA^{\nu b}(x))\nonumber \\
&&-\gamma^\nu ({\overleftarrow \partial}^\lambda +igT^bA^{\lambda b}(x)) -\gamma^\lambda  ({\overleftarrow \partial}^\nu +igT^bA^{\nu b}(x))  ] \psi(x).
\label{yakxtz}
\eea
From eq. (\ref{yakxtz}) we find
\bea
\partial_\nu J^{\nu \lambda \delta}(x)=\partial_\nu [T_S^{\nu \delta}(x) x^\lambda - T_S^{\nu \lambda}(x) x^\delta]=0
\label{yfinem}
\eea
which shows that the symmetric and gauge invariant energy-momentum tensor $T_S^{\nu \lambda}(x)$ of the the quark plus the Yang-Mills field in Yang-Mills theory in eq. (\ref{yakxtz}) can be obtained from the gauge invariant Noether's theorem by using combined Lorentz transformation plus local non-abelian gauge transformation without requiring Belinfante-Rosenfeld tensor.

Hence we find from eqs. (\ref{yakxtz}) and (\ref{yfinem}) that the Belinfante-Rosenfeld tensor is not necessary in Yang-Mills theory to obtain the symmetric and gauge invariant energy-momentum tensor of the the quark plus the Yang-Mills field in Yang-Mills theory.

\section{Non-Vanishing Boundary Surface Term and Non-Existence of Gauge Invariant Spin and Orbital Angular Momentum of Yang-Mills Field }

Eq. (\ref{yovbtysa}) is the gauge invariant definition of the angular momentum of the quark plus the Yang-Mills field in Yang-Mills theory derived from the first principle, {\it i. e.}, by using combined Lorentz transformation plus local non-abelian gauge transformation. Using eq. (\ref{ydelt}) in (\ref{yovbtysa}) we find
\bea
&& {\vec J}_{q+g}= {\vec J}_{g}+{\vec L}_q+{\vec S}_q= \int d^3x~[ {\vec r} \times [{\vec E}^b(x) \times {\vec B}^b(x)]+ {\vec r} \times [\psi^\dagger(x) [-i{\overrightarrow {\vec D}}+i{\overleftarrow {\vec D}}] \psi(x)]+\psi^\dagger(x) ~{\vec \Sigma}~ \psi(x)]\nonumber \\
&&=\int d^3x~ [{\vec E}^b(x)\times {\vec A}^b(x)+{ E}^{jb}(x) ~{\vec r} \times {\vec \nabla} A^{jb}(x) + {\vec r} \times [\psi^\dagger(x) [-i{\overrightarrow {\vec \partial}}+i{\overleftarrow {\vec \partial}}] \psi(x)]+\psi^\dagger(x) ~{\vec \Sigma}~ \psi(x)] \nonumber \\
&&- \int d^3x~ \partial_j [{ E}^{jb}(x) ~{\vec r} \times {\vec A}^b(x)].
\label{yovbt}
\eea
In Yang-Mills theory the Yang-Mills potential (the color potential) $A^{\nu b}(x)$ produced at $x^\nu$ by the quark in motion at $X^\nu(\tau)$ with four-velocity $u^\nu(\tau)=\frac{dX^\nu(\tau)}{d\tau}$ is given by \cite{ynj1,yne1}
\bea
&&A_\nu^b(x) =  \frac{u_\nu(\tau_0)}{s} q^c(\tau_0)[\frac{e^{g\int ds \frac{Q(\tau_0)}{s}}-1}{g\int ds \frac{Q(\tau_0)}{s}}]_{bc},~~~~~~~~~s=u(\tau_0)\cdot (x-X(\tau_0)),\nonumber \\
&&Q^{bc}(\tau_0)=f^{bcd}q^d(\tau_0),~~~~~(x-X_0(\tau_0))^2=0
\label{ylwp}
\eea
where $q^b(t)$ is the time dependent fundamental color charge of the quark. Similarly the SU(3) pure gauge potential $A^{\nu b}_{pure}(x)$ in Yang-Mills theory is given by \cite{ynj1,yne1}
\bea
A^{\nu b}_{pure}(x) =  \frac{\beta^\nu_c}{ s_c} q^d(\tau_0)[\frac{e^{g\int ds_c \frac{Q(\tau_0)}{s_c}}-1}{g\int ds_c \frac{Q(\tau_0)}{s_c}}]_{bd},~~~~~~~\beta^2_c=0,
~~~~~~~~~s_c=\beta_c \cdot (x-X(\tau_0)).
\label{ylwpg}
\eea
Because of the form in eq. (\ref{ylwp}) [which, unlike Maxwell theory, is not inversely proportional to distance] we find the non-vanishing boundary surface term
\bea
\int d^3x~ \partial_j [{ E}^{jb}(x) ~{\vec r} \times {\vec A}^b(x)] \neq 0
\label{ypvbt}
\eea
in Yang-Mills theory which implies that for angular momentum the boundary surface term in eq. (\ref{yovbt}) does not vanish in the Yang-Mills theory.

From eqs. (\ref{ypvbt}) and (\ref{yovbt}) we find
\bea
&& {\vec J}_{q+g}= {\vec J}_{g}+{\vec L}_q+{\vec S}_q=  \int d^3x~[ {\vec r} \times [{\vec E}^b(x) \times {\vec B}^b(x)]+ {\vec r} \times [\psi^\dagger(x) [-i{\overrightarrow {\vec D}}+i{\overleftarrow {\vec D}}] \psi(x)]+\psi^\dagger(x) ~{\vec \Sigma}~ \psi(x)]\nonumber \\
&&\neq \int d^3x~ [{\vec E}^b(x)\times {\vec A}^b(x)+{ E}^{jb}(x) ~{\vec r} \times {\vec \nabla} A^{jb}(x) + {\vec r} \times [\psi^\dagger(x) [-i{\overrightarrow {\vec \partial}}+i{\overleftarrow {\vec \partial}}] \psi(x)]+\psi^\dagger(x) ~{\vec \Sigma}~ \psi(x)]\nonumber \\
\label{yqvbt}
\eea
which gives
\bea
&& {\vec J}_{q+g}= {\vec L}_q+{\vec S}_q+{\vec J}_{g}\neq {\vec L}^{non-inv}_q+{\vec S}_q+{\vec L}^{non-inv}_{g}+{\vec S}^{non-inv}_{g}
\label{yfingt}
\eea
where ${\vec J}_{q+g}$ is the gauge invariant angular momentum of the quark plus the Yang-Mills field as given by eq. (\ref{yovbtysa}), ${\vec J}_{g}$ is the gauge invariant angular momentum of the Yang-Mills field as given by eq. (\ref{yohav}), ${\vec L}_{q}$ is the gauge invariant orbital angular momentum of the quark in Yang-Mills theory as given by eq. (\ref{yoquakh}), ${\vec S}_{q}$ is the gauge invariant spin angular momentum of the quark in Yang-Mills theory as given by eq. (\ref{yoqua}), ${\vec L}^{non-inv}_{g}$ is the gauge non-invariant orbital angular momentum of the Yang-Mills field as given by eq. (\ref{yhjgk}), ${\vec S}^{non-inv}_{g}$ is the gauge non-invariant spin angular momentum of the Yang-Mills field as given by eq. (\ref{ydspn}) and ${\vec L}^{non-inv}_{q}$ is the gauge non-invariant orbital angular momentum of the quark in Yang-Mills theory as given by eq. (\ref{yahjgka}).

Hence, as opposed to various gauge invariant (and gauge non-invariant) definition of the spin angular momentum of the Yang-Mills field in the literature \cite{yjaffe1,yji1,ywata,yhata,ygold}, we find from eq. (\ref{yfingt}) that the notion of the gauge invariant definition of the orbital angular momentum ${\vec L}_g$ of the Yang-Mills field and the notion of the gauge invariant definition of the spin angular momentum ${\vec S}_g$ of the Yang-Mills field do not exist from the first principle in Yang-Mills theory because of the non-vanishing boundary surface term [see eq. (\ref{ypvbt})] in Yang-Mills theory although the definition of the gauge invariant orbital angular momentum ${\vec L}_q$ of the quark and the definition of the gauge invariant spin angular momentum ${\vec S}_q$ of the quark exist from the first principle in Yang-Mills theory.

\section{conclusions}
The gauge invariant definition of the spin dependent gluon distribution function from first principle is necessary to study the proton spin crisis at high energy colliders. In this paper we have derived the gauge invariant Noether's theorem in Yang-Mills theory by using combined Lorentz transformation plus local non-abelian gauge transformation. We have found that the definition of the gauge invariant spin (or orbital) angular momentum of the Yang-Mills field does not exist in Yang-Mills theory although the definition of the gauge invariant spin (or orbital) angular momentum of the quark exists.

We have shown that the gauge invariant definition of the spin angular momentum of the Yang-Mills field in the literature  \cite{yjaffe1,yji1,ywata,yhata,ygold} is not correct because of the non-vanishing boundary surface term in Yang-Mills theory. We have also found that the Belinfante-Rosenfeld tensor in Yang-Mills theory is not required to obtain the symmetric and gauge invariant energy-momentum tensor of the quark and the Yang-Mills field.

Note that in addition to heavy-ion collisions at RHIC to study quark-gluon plasma \cite{ynka1,ynka2,ynka3} the RHIC experiment involves polarized proton-proton collisions at high energy to study proton spin physics \cite{yrhic,yrhic1} and to extract spin dependent gluon distribution function inside the proton. From this point of view the study of the gauge invariant definition of the spin dependent gluon distribution function at high energy coliders is necessary, especially when we have shown in this paper that we do not have a gauge invariant definition of the spin dependent gluon distribution function inside the proton in QCD at high energy colliders from the first principle.

\end{document}